\documentclass[%
 reprint,twocolumn,
 superscriptaddress,
 nobibnotes,
 amsmath,amssymb,
 prl,
floatfix,
longbibliography,
]{revtex4-1}
\usepackage[usenames,dvipsnames,svgnames,table]{xcolor}
\colorlet{m}{Magenta}
\colorlet{k}{Black}

\usepackage{graphicx}
\usepackage[outdir=./]{epstopdf}

\usepackage{dcolumn}
\usepackage{bm}
\usepackage[colorlinks=true,citecolor=blue]{hyperref}
\usepackage[rightcaption]{sidecap}
\usepackage[utf8]{inputenc}

\begin{document}

\title{Fighting Broken Symmetry by Doping: Toward Polar Resonant Tunneling Diodes with Symmetric Characteristics}

\author{Jimy Encomendero}
\email{jje64@cornell.edu}
\affiliation{\hbox{School of Electrical and Computer Engineering, Cornell University, Ithaca, New York 14853 USA}}%

\author{Vladimir Protasenko}
\affiliation{\hbox{School of Electrical and Computer Engineering, Cornell University, Ithaca, New York 14853 USA}}%

\author{Farhan Rana}
\affiliation{\hbox{School of Electrical and Computer Engineering, Cornell University, Ithaca, New York 14853 USA}}%

\author{Debdeep Jena}
\email{djena@cornell.edu}
\affiliation{\hbox{School of Electrical and Computer Engineering, Cornell University, Ithaca, New York 14853 USA}}%
\affiliation{\hbox{Department of Materials Science and Engineering, Cornell University, Ithaca, New York 14853 USA}}%
\affiliation{\hbox{Kavli Institute at Cornell for Nanoscale Science, Cornell University, Ithaca, New York 14853, USA}}

\author{Huili Grace Xing}
\email{grace.xing@cornell.edu}
\affiliation{\hbox{School of Electrical and Computer Engineering, Cornell University, Ithaca, New York 14853 USA}}%
\affiliation{\hbox{Department of Materials Science and Engineering, Cornell University, Ithaca, New York 14853 USA}}%
\affiliation{\hbox{Kavli Institute at Cornell for Nanoscale Science, Cornell University, Ithaca, New York 14853, USA}}%

\begin{abstract}
The recent demonstration of resonant tunneling transport in nitride semiconductors has led to an invigorated effort to harness this quantum transport regime for practical applications. In polar semiconductors, however, the interplay between fixed polarization charges and mobile free carriers, results in highly asymmetric tunneling transport characteristics. Here, we investigate the possibility of using degenerately doped contact layers to screen the built-in polarization fields and recover symmetric resonant injection. Thanks to a high doping density, negative differential conductance is observed under both bias polarities of GaN/AlN resonant tunneling diodes (RTDs). Moreover, our analytical model reveals a lower bound for the minimum resonant-tunneling voltage achieved via uniform doping, owing to the dopant solubility limit. Charge storage dynamics is also studied by impedance measurements, showing that at close-to-equilibrium conditions, polar RTDs behave effectively as parallel-plate capacitors. These mechanisms are completely reproduced by our model, providing a theoretical framework useful in the design and analysis of polar resonant-tunneling devices.
\\
\end{abstract}

\maketitle

\section{Introduction}

III-nitride semiconductor heterostructures stand out as a highly versatile platform for tailoring electronic levels via quantum confinement because of their wide range of achievable bandgaps (0.7 eV--6.2 eV). This important property, coupled with their high breakdown electric fields [\onlinecite{Maeda2018}], high thermal conductivity [\onlinecite{Mion2006}], and high electron saturation velocity [\onlinecite{Bajaj2015}], make III-nitride semiconductors an unmatched material system for the development of high-power electronic and photonic devices. 

From ultrafast electronic oscillators [\onlinecite{Feiginov2019}] to room-temperature intersubband lasers [\onlinecite{Terashima2011}], and plasma-wave amplifiers [\onlinecite{Sensale2013,Condori2016}], III-nitride materials hold the promise for the development of compact, high-power terahertz sources. These new functionalities are enabled by the broadband gain provided by the resonant tunneling phenomena which can be employed not only to counteract lossy circuit elements but also provide electronic and optical gain within the terahertz band [\onlinecite{Feiginov2019,Kohler2002}]. The synergy between the ultrafast tunneling transport and the outstanding power capabilities of III-nitride semiconductors represents a tantalizing opportunity for the development of nitride-based terahertz optoelectronics.

Even though research on nitride resonant tunneling transport started almost two decades ago [\onlinecite{Kikuchi2001,Kikuchi2002}], in the past three years important breakthroughs have been achieved due to recent advances in epitaxial growth, polar heterostructure design, and device fabrication [\onlinecite{Encomendero2020APS,Encomendero2016X,Growden2016,Encomendero2017}]. In particular, the demonstration of reliable room-temperature negative differential conductance (NDC) [\onlinecite{Sakr2012,Encomendero2016X,Growden2016,Encomendero2017,Wang2018,Growden2018,Encomendero2020}] and the first nitride-based resonant tunneling oscillator [\onlinecite{Encomendero2018,Xing2019}] have led to a better understanding of resonant tunneling physics through polar semiconductors. These milestones have reignited interest in the development of practical device applications enabled by nitride resonant tunneling injection.

Because of their noncentrosymmetric crystal structure, nitride heterostructures grown along the polar axes exhibit spontaneous and piezoelectric polarization charges localized at the heterointerfaces [\onlinecite{Ambacher2000}]. These interfacial sheets of charge are of central importance in the understanding of resonant tunneling transport through polar semiconductors. Their effects have been recently elucidated by the introduction of a unified theoretical framework, which explains all the features experimentally observed in the current-voltage characteristics of GaN/AlN resonant tunneling diodes (RTDs) [\onlinecite{Encomendero2019}].

The presence of the interfacial polarization charges ($\pm q\sigma_\pi$), and associated net polarization fields ($F_\pi=\pm q\sigma_\pi/\epsilon_s$), generate intense internal electric fields which in turn redistribute free carriers across the double-barrier active region and surrounding contacts of the heterojunction. Here, $q$ is the absolute value of the electron charge, and $\sigma_\pi$ is the polarization charge density at the GaN/AlN heterojunction. Under equilibrium conditions, electrons are confined within the emitter accumulation well, on the substrate side; whereas on the collector contact, a depletion region is formed [see Fig.~1(b)]. With typical contact doping concentrations $N_d\sim 1\times 10^{19}$~cm$^{-3}$ and depletion widths $t_d\sim 10$~nm, the collector space-charge density $N_d\times t_d= 1\times 10^{13}$~cm$^{-2}$, is of the same order as the polarization charge density $\sigma_\pi\sim 5\times10^{13}$~cm$^{-2}$. This observation reveals that space-charge effects due to contact doping play an important role in the electric field distribution and conduction band profile of polar double-barrier heterostructures. However the important consequences of the interplay between the fixed polarization charges in the active region and mobile charge in the access regions have not been throughly discussed in the literature.

The purpose of the present study is to experimentally determine the role played by the space-charge regions on the resonant tunneling characteristics of polar RTDs. By systematically varying the extension of the space-charge layer, we show an enhanced control over the energies of the bound states, thereby reducing the resonant tunneling voltage. With the use of theoretical calculations we assess the device performance as the width of the space-charge region is modulated via impurity doping. With this approach we show that polar RTDs exhibit a lower bound for the minimum achievable resonant-tunneling voltage as the doping density approaches the solubility limit [\onlinecite{Faria2012,Lugani2014}].

Furthermore, cryogenic transport measurements reveal repeatable NDC under both bias polarities. These findings shed light onto the important effects of the space-charge regions in the transmission amplitude of the resonant tunneling current. This connection manifests itself not only in the dc characteristics, but also in the small-signal frequency response of the diodes. After measuring the admittance of these devices, we conclude that under close-to-equilibrium conditions, polar RTDs behave effectively as parallel-plate capacitors. This phenomenon, completely captured by our recently developed polar RTD model [\onlinecite{Encomendero2017}], allows us to derive an analytical expression for the bias-dependent RTD capacitance. The good quantitative agreement between the theoretical and measured capacitances, confirms the validity of our model, providing a clear physical picture for the charge storage mechanism of polar RTDs.

\begin{figure}[t]
	\centering
	\includegraphics[width=0.5\textwidth]{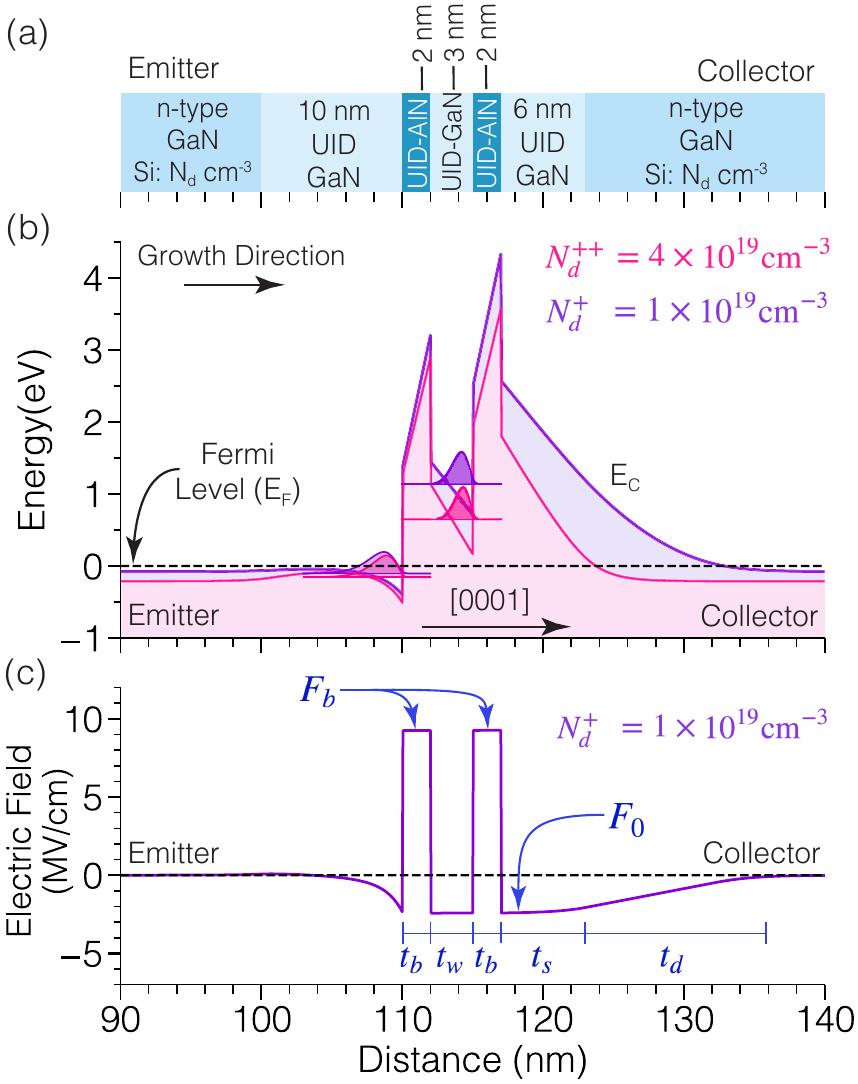}
	\hypertarget{Fig-RTD}{}
	\caption{GaN/AlN double-barrier RTD heterostructures. (a) Schematic representation of the complete device structure grown by molecular beam epitaxy on single-crystal \textit{n}-type GaN substrates. Two device designs with varying doping concentration $N_d$ are synthesized and fabricated. (b) The equilibrium conduction-band energy profile of each sample is calculated by solving Poisson and Schr\"odinger equations self-consistently. As a result of the redistribution of free carriers, the conduction band profile and internal electric fields are modulated resulting in a shift in the energies of the resonant levels inside the well. (c) The electric field profile for the RTD structure featuring a doping density $N_d^+=1\times 10^{19}$~cm$^{-3}$ is also obtained from numerical calculations. The various layer thicknesses are also indicated: $t_d$ is the extension of the collector depletion region, $t_s$ is the thickness of the collector spacer, and $t_b$ and $t_w$ represent the thickness of the tunneling barriers and width of the quantum well, respectively.}
\end{figure}

\section{G\lowercase{a}N/A\lowercase{l}N RTD Electrostatics}

The prototypical epitaxial structure of a GaN/AlN RTD is schematically depicted in Fig.~1(a). Heavily doped $\textit{n}$-type GaN contacts are employed to inject electrons across the double-barrier structure which consists of unintentionally (UID) doped GaN spacers next to each contact, two symmetric AlN tunneling barriers, and a GaN-quantum well hosting the resonant states. Under equilibrium conditions, the active region exhibits a peculiar conduction band profile due to the interfacial polarization charges ($\pm q\sigma_\pi$) and space-charge distribution in the contacts. The positive charge $+q\sigma_\pi$ localized at the interface between the emitter UID spacer and AlN barrier, attracts free electrons that gather in the accumulation well next to this heterointerface [See Fig.~1(b)]. In contrast, on the collector region, the negative polarization sheet of charge at the AlN/GaN heterointerface repels electrons, resulting in a wide depletion region. As can be inferred from Figs.~1(b) and (c), the width of the depletion layer $t_d$, determines the maximum electric field $F_0$ on the collector spacer, which can be calculated using the depletion approximation [\onlinecite{Encomendero2017}]. Therefore, the magnitude of $F_0$ can be expressed as:
\begin{equation}
F_0=qN_dt_d/\epsilon_s,
\end{equation}
where $N_d$ is the doping concentration in the GaN contacts, and $\epsilon_s$ is the semiconductor dielectric constant. It is helpful to point out that the extension of the depletion region $t_d$, depends on the voltage bias applied to the double-barrier structure [\onlinecite{Encomendero2017}]. Consequently under non-equilibrium conditions, both $t_d$ and $F_0$ would be modulated.

Since no free charge is present across the collector UID-GaN spacer, the electric field within this layer remains constant at $F_0$. The negative sheet of polarization charge present at the interface between the UID-layer and collector barrier flips the direction of the electric field present inside the collector tunneling barrier [See Fig.~1(c)]. Considering the fact that the sheets of polarization charge present on each side of the collector AlN barrier are of opposite sign, no net charge exists between the collector space-charge region and quantum well. Therefore, the electric polarization of the GaN well is equivalent to the maximum electric field within the collector depletion region, and thus $F_w=F_0$. This connection between the space-charge electric field and the electrostatic polarization of the barriers and quantum well reveals the influence of the collector doping on the band profile of the active region and in turn determines the energies of the resonant levels at any given bias condition.

To experimentally study these effects, we prepare GaN/AlN double-barrier heterostructures by molecular beam epitaxy (MBE). The epitaxial layers are grown on the c-plane of single-crystal \textit{n}-type GaN substrates, maintaining a constant growth rate of 3~nm/minute [\onlinecite{Encomendero2019}]. The substrate temperature, measured by a thermocuople, is kept constant at 750~$^\circ$C while maintaining metal-rich conditions and step-flow growth mode throughout the whole epitaxial process [\onlinecite{Encomendero2017}].

Epitaxy begins with the growth of a highly-doped \textit{n}-type GaN emitter contact layer, extending 100~nm with a nominal silicon concentration $N_d$~cm$^{-3}$. As shown in Fig.~1(b), two different RTD designs were prepared with different silicon doping concentrations: $N_d^+=1\times 10^{19}$~cm$^{-3}$, and $N_d^{++}=4\times 10^{19}$~cm$^{-3}$. The doping level, controlled by the temperature of the silicon effusion cell, was previously calibrated employing a separate sample in which a series of $\textit{n}$-type GaN layers were grown with varying doping levels. Using secondary ion mass spectroscopy we determine the correspondence between the donor concentration and effusion cell temperature employed during the RTD growth. To minimize electronic decoherence due to impurity scattering, a 10-nm UID GaN spacer is grown next to the emitter GaN contact, separating it from the active structure. The double-barrier active region consists of two symmetric AlN tunneling barriers, extending 2~nm, which confine the resonant states within a 3-nm-GaN quantum well. To limit the extension of the collector depletion region, we employ an RTD design featuring asymmetric spacers; therefore, the thickness of the UID collector spacer is 6~nm. Finally, a 100-nm $\textit{n}$-type GaN layer, acting as collector contact, completes the device structure. as shown in Fig.~1(a). After growth, atomic force microscopy (AFM) is employed to measure the topography of the as-grown samples, revealing clear atomic steps across scanned areas of $20\times 20$~$\mu $m$^2$ with sub-nanometer rms surface roughness.

Figure~1(b) displays the equilibrium energy band diagrams of the grown samples calculated with a self-consistent Schr\"odinger-Poisson solver [\onlinecite{Tan1990}]. We can see that by increasing the doping concentration, the energy of the ground-state $E_1$ reduces by $\Delta E_1=-488$~meV. This modulation is the result of a combination of effects attributed to: (a) the tuning of the Fermi energy within the degenerately-doped contact layers, (b) the reduction of the electrostatic polarization within the tunneling barriers, and (c) the stronger quantum-confined Stark effect (QCSE) inside the well. We can quantify each of these effects to understand their individual influence over the energies of the resonant levels:

(a) Owing to the high concentration of silicon donors, the \textit{n}-type GaN contact layers become degenerately doped. This behavior has been verified with both numerical self-consistent calculations [\onlinecite{Tan1990}] and the analytical expression we owe to Joyce and Nixon [\onlinecite{Joyce1977}]. Consequently, as the doping level increases, the bottom of the conduction band shifts towards lower values with respect to the Fermi level which we choose as our reference [See Fig. 1(b)]. We can obtain a first-order approximation for this energy shift employing the Fermi gas model [\onlinecite{Ashcroft1976}]. Using this approach, we can express $E_c$ as a function of the doping concentration: $E_f-E_c\approx\frac{\hbar^2}{2m^\star}\left(3\pi^2N_d\right)^{2/3}$, where $\hbar$ is the reduced Planck constant and $m^\star$ is the GaN effective mass. Consequently, the whole band profile of the heterostructure shifts with respect to the Fermi level by a factor $\Delta E_{E_c}=-\frac{3^{2/3}\pi^{4/3}\hbar^2}{2m^*}\left[\left(N_d^{++}\right)^{2/3}-\left(N_d^+\right)^{2/3}\right]$. 

(b) From Fig.~1(c), we can see that there exists a direct connection between the magnitude of electric field inside the collector region ($F_0$), the interface polarization charge ($-q\sigma_\pi$), and the electrostatic polarization of the barriers ($F_b$) which can be written as $F_b-q\sigma_\pi/\epsilon_s=F_0$. Consequently, when the collector electric field becomes more negative, it leads to a concomitant reduction in the magnitude of the field inside the tunneling barriers. This modulation results in a shift of the energy profile of the well, moving it closer to the Fermi level by a factor $\Delta E_{F_b}=-q\Delta F_b t_b=-q\Delta F_0 t_b$; where $\Delta F_0$ is the variation of the maximum electric field inside the space-charge region [See Eq. (1) and Fig.~1(c)], and $t_b$ is the barrier thickness.

(c) Finally the intense electrostatic polarization of the quantum well results in the modulation of the resonant energies due to QCSE. An approximate value for this energy shift $\Delta E_{F_w}$, can be calculated within the framework of perturbation theory [\onlinecite{Sakurai2011}]. Thus, we find that $\Delta E_{F_w}=-q\Delta F_wt_w/2=-q\Delta F_0t_w/2$, where $t_w$ is the width of the GaN well as depicted in Fig.~2(c).

\begin{SCfigure*}
	\includegraphics[width=0.6828\textwidth]{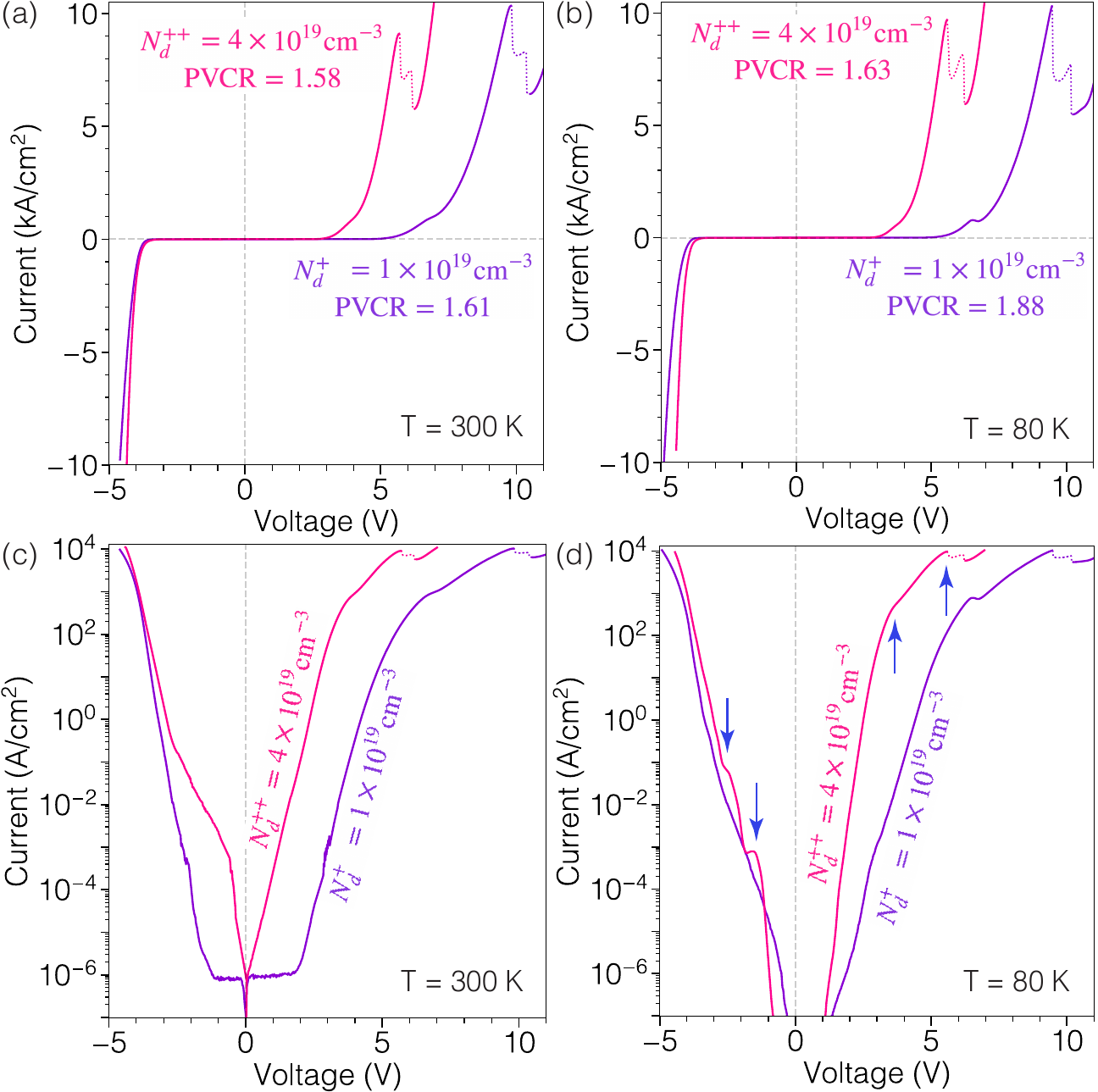}
	\caption{Current-voltage (\textit{J-V}) characteristics of GaN/AlN resonant tunneling diode (RTDs) with different contact doping concentrations measured at cryogenic and room temperatures. (a),(b) When the doping level at the contacts is increased from $N_d^+=1\times 10^{19}$~cm$^{-3}$ to $N_d^{++}=4\times 10^{19}$~cm$^{-3}$, the energy of the lowest resonant level is reduced, resulting in a lower peak voltage under forward bias. Within the region of NDC, plotted with dotted lines, parasitic oscillations are measured due to the RTD gain, which produces the chair-like feature. The peak and valley currents are the maximum and minimum currents that delimit the NDC region. (b) Under cryogenic conditions, the reduction of the thermal current component results in an increased peak-to-valley current ratio (PVCR) which is measured at 1.88 (1.63) for the RTD structure with the lower (higher) doping level. (c),(d) Logarithmic plots of the same data displayed in (a),(b), which reveal the exponential modulation of the tunneling current under non-equilibrium conditions. (d) Multiple resonant tunneling features, indicated by the arrows, are observed under forward and reverse bias. The origin of these features is explained in Fig.~3.}
\end{SCfigure*}

With the previous analysis in mind, we can calculate each of the partial contributions to the energy shift of the resonant levels. Theoretical calculations indicate that $\Delta F_0=0.9$~MV/cm, when the doping increases from $N_d^+=1\times 10^{19}$~cm$^{-3}$ to $N_d^{++}=4\times 10^{19}$~cm$^{-3}$. Thus, we find that $\Delta E_{F_b}=-180$~meV, $\Delta E_{F_w}=-135$~meV, and $\Delta E_{E_c}=-129$~meV. Therefore the total energy shift given by $\Delta E_1=\Delta E_{E_c}+\Delta E_{F_b}+\Delta E_{F_w}$, is $\Delta E_1=-444$~meV; in reasonable agreement with $-488$~meV, obtained from self-consistent calculations. This detailed analysis allows us to fully appreciate the implications of contact doping and space-charge effects in the electrostatics and resonant tunneling conditions of polar RTDs. As we will see in the following sections, these important effects are experimentally observed in multiple GaN/AlN RTD designs with different doping concentrations, thus providing valuable insight into the dc and ac operation of polar resonant tunneling devices.

\section{Space-Charge Effects on Resonant Tunneling Transport}

To study charge transport, we fabricate RTDs employing a self-aligned process described in our previous work [\onlinecite{Encomendero2017}]. Devices with areas between $6\times 6$~$\mu m^2$ and $20\times 20$~$\mu m^2$ were fabricated and tested under both bias polarities showing repeatable current-voltage characteristics and NDC. We point out that forward current injection corresponds to the electronic flow from the substrate to the collector contact. Typical current-voltage (\textit{J-V}) characteristics, measured from devices with areas of $6\times6$~$\mu$m$^2$, are displayed in Fig. 2. As can be seen from Figs. 2(a) and (b), when the doping density increases, the resonant tunneling voltage transitions from 9.6~V to 6.6~V. This effect is a direct consequence of the shift in ground-state energy which was quantitatively analyzed in the preceding section.

To suppress the thermal component of the valley current we tested our devices at cryogenic conditions. Figure~2(b) shows the $\textit{J-V}$ curves measured at $T=80$~K, resulting in a peak-to-valley current ratio (PVCR) of 1.88 (1.63) for the RTD structure featuring the lower (higher) doping concentration. Whereas the peak voltage under forward bias is strongly dependent on the dopant concentration, the reverse-bias threshold voltage, measured at $V_\text{th}=-4.2\pm 0.2$~volts, is independent of doping [See Figs.~2(a) and (b)]. This result is consistent with our theoretical model for polar RTDs, from which we established that the threshold voltage depends only on the barrier thickness and magnitude of the internal polarization fields, and not on donor concentration [\onlinecite{Encomendero2017}].

The broken symmetry in the tunneling transport characteristics of GaN/AlN RTDs stems from the intense electrostatic polarization of the double-barrier active region which results in a redistribution of free carriers. Owing to this internal charge segregation, the profile of the tunneling barriers becomes highly asymmetric, resulting in single-barrier transmission coefficients which differ by several orders of magnitude [\onlinecite{Encomendero2019}]. When a voltage is applied across the active region, the electronic tunneling transmission and therefore the RTD conductance are both exponentially modulated. This exponential increase in tunneling current can be clearly seen in Figs.~2(c) and (d), which show the \textit{J-V} curves in logarithmic scale. Under cryogenic conditions, multiple resonant tunneling signatures, indicated by the arrows, can be identified in Fig.~2(d). These features correspond to the energy alignment between the two- and three-dimensional pockets of electrons present in the emitter side, and the resonant tunneling subbands confined within the double-barrier structure. To get further insight into the charge transport mechanisms responsible for these features, we examine the RTD differential conductance.

The semilogarithmic plot in Fig.~3(a) displays the exponential modulation of the differential conductance for both RTD designs operating at cryogenic conditions. Under reverse bias we can identify two peaks, indicated by the arrows, which correspond to the alignment between the collector Fermi level ($E_f^C$) and the resonant levels inside the quantum well. The conductance peak labeled by $E_f^C\rightarrow E_1$ originates from resonant injection into the ground-state and exhibits repeatable NDC, with a minimum value of $-6\times 10^{-4}$~S/cm$^2$. These findings constitute the first report of NDC measured from GaN/AlN RTDs under reverse bias operation. It should be noted however that reverse-bias conductance oscillations were documented at cryogenic [\onlinecite{Wang2018}], and room temperatures [\onlinecite{Encomendero2019}]. The possibility of resonant tunneling injection, manifested in reverse-bias NDC, was anticipated by previous theoretical considerations [\onlinecite{Sakr2011}]. Under these conditions, the marked asymmetry in the transparency of the barriers leads to a highly attenuated resonant tunneling transmission [\onlinecite{Encomendero2019}]. In this sense, minimizing incoherent charge injection is of great importance to allow resonant tunneling become the dominant transport mechanism within this bias regime.

Resonant tunneling into the first excited state is also identified from the conductance peak at $V_\text{bias}=-2.4$~V [See Fig.~3(a)]. The RTD band diagram for this resonant configuration is displayed in Fig.~3(c), which reveals a significant asymmetry between the single-barrier tunneling distances. This imbalance in single-barrier transmission probabilities, experienced by the electrons, leads to a strongly attenuated resonant tunneling transmission.  Despite these effects, the resonant conductance peak is clearly visible in the conductance-voltage ($\textit{G-V}$) curve for the RTD featuring the highest doping concentration. These findings constitute the first experimental demonstration of reverse-bias resonant tunneling injection into the ground-state as well as the first-excited state. Therefore, this important milestone open up the possibility of harnessing the reverse-bias transport regime for the engineering of population inversion in nitride-based quantum cascade structures [\onlinecite{Berland2011,Martin2011}].

\begin{SCfigure*}
	\hypertarget{Fig-GV}{}
	\includegraphics[width=0.71\textwidth]{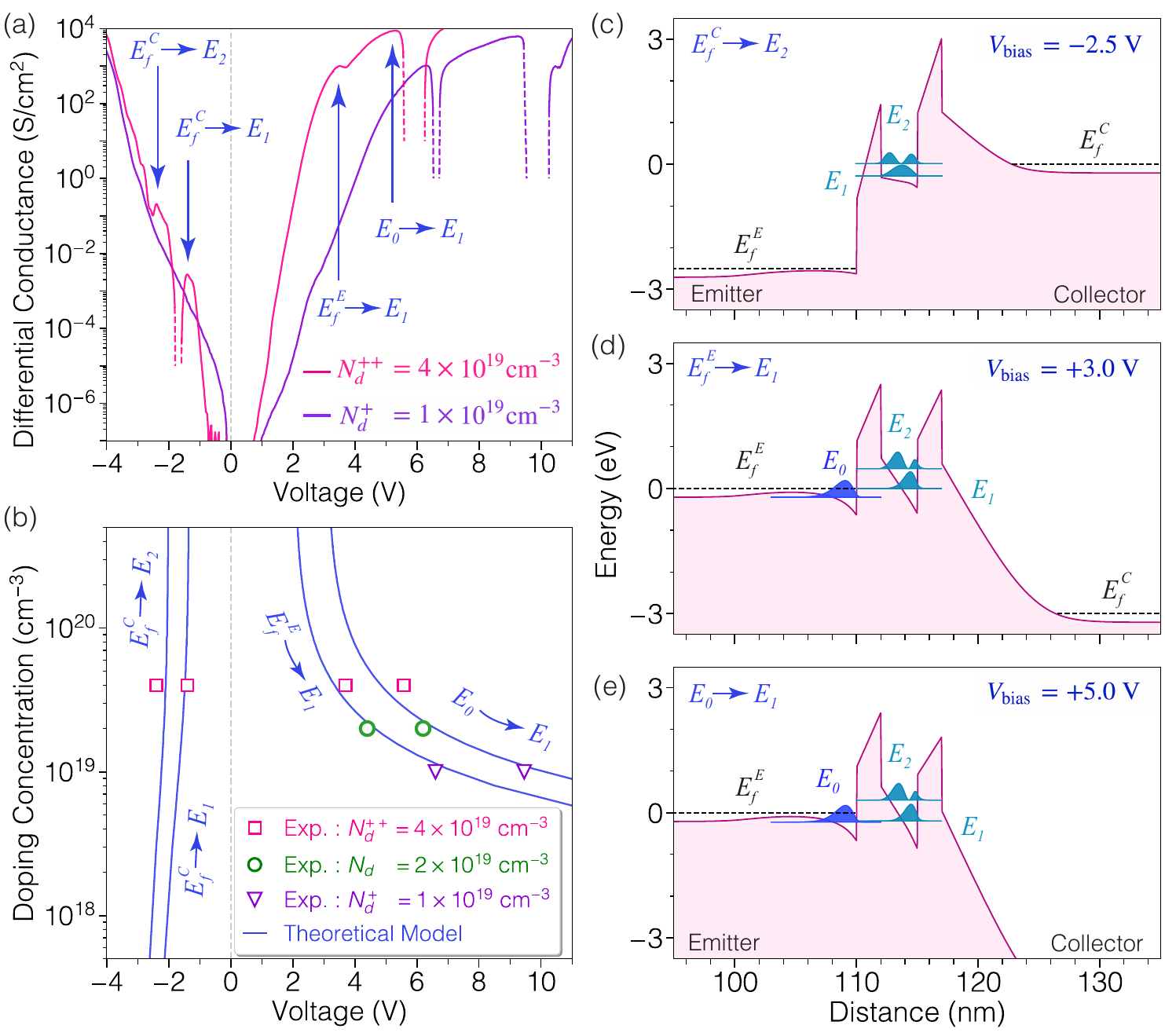}
	\caption{Under cryogenic temperatures ($T=80$~K), various resonant tunneling features are observed in the current-voltage ($\textit{J-V}$) characteristics of GaN/AlN resonant tunneling diodes (RTDs). (a) The differential conductance reveals multiple peaks, indicated by the arrows, which correspond to the resonant alignment between the electron reservoirs in the contacts and the resonant levels within the well. (b) The voltages required to attain these resonances are strongly dependent on the doping concentration at the contacts. The experimentally measured resonant voltages are displayed as a function of the doping concentration, showing a good quantitative agreement with theoretical calculations. The experimental data points shown as circles ($\circ$) were taken from Ref. [\onlinecite{Encomendero2019}]. (c)-(e) To elucidate the origin of the conductance peaks, we calculate the RTD band diagrams for each of the resonant configurations for $N_d^{++}=4\times10^{19}$~cm$^{-3}$, under forward- and reverse-bias conditions.}
\end{SCfigure*}

Within the forward bias regime, we can identify two peaks in the differential conductance, indicated by the arrows in Fig.~3(a). These features correspond to different injection mechanisms due to the dual nature of carriers present in the emitter. We associate the transition labeled by $E_f^E\rightarrow E_1$, with the injection of carriers from the continuum of states, forming the emitter Fermi sea, into the ground-state of the GaN well. The onset of this resonant condition is shown in Fig.~3(d), revealing the alignment between emitter Fermi level $E_f^E$, and the lowest subband in the well with energy $E_1$. This tunneling feature, occurring at a voltage lower than the main resonant peak, has been reported previously in arsenide [\onlinecite{Eaves1988,Wu1990,Buchanan1990,Koenig1990,Wu1991,Wosinski1998}] and nitride RTDs [\onlinecite{Encomendero2017,Wang2018}]. In our devices, the injection of 3D-carriers into the well peaks at $V_\text{bias}=+3.5$~V and $V_\text{bias}=+6.6$~V for the $N_d^{++}$-- and $N_d^+$--RTD, respectively.

As the applied bias increases, the ground-state $E_1$ is pulled bellow the emitter Fermi level and becomes degenerate with the energy $E_0$ of the accumulation subband [See Fig.~3(e)]. In this regime, the RTD conductance is dominated by the resonant-tunneling path established by the coupling of these subbands. It should be noted that the confinement introduced by the accumulation well, results in a lower bound for the minimum energy of the electrons injected into the well, given by $E_0$. Therefore, the detuning from this resonant configuration reduces the amount of carriers that tunnel coherently into the resonant level, which manifests in the characteristic NDC of the main resonant peak shown in Fig.~2(b).

The voltage required to attain each of the resonant configurations indicated by the arrows in Fig.~3(a) depends on the doping density in the contacts. The tuning of these resonant voltages as a function of the doping concentration can be determined by self-consistent calculations of the biased device structure [see for example Figs.~3(c)-(e)]. Figure 3(b) shows the theoretical values of the resonant voltages as the doping concentration is modulated between $N_d=5\times 10^{17}$ and $N_d=5\times 10^{20}$~cm$^{-3}$. These numerical results are consistent with the experimentally measured resonant-tunneling features, which are also included in the same figure for comparison. From this plot, we notice the considerable asymmetry in the resonant tunneling voltages with respect to the bias polarity. This asymmetric behavior manifests not only in the magnitude of the resonant voltages but also in its dependence on the doping concentration. Whereas reverse-bias resonances are almost independent of the doping level, forward resonant tunneling voltages exhibit a strong dependence on this design parameter. These trends originate from the relationship between the contact doping levels and width of the depletion layer which determines the fraction of applied voltage that is dropped across this region. Under forward injection the depletion layer widens, resulting in a larger resonant-tunneling voltage which depends strongly on the doping level, as can be seen in Fig.~3(b). In contrast, under reverse bias operation, the collector depletion region shrinks, thus leading to a reverse-bias resonant voltage that is almost independent of the doping level. The invariance of the reverse-bias resonant condition might be helpful for the design of quantum cascade structures in which resonant injection would be robust against potential variations in the doping level, in stark contrast to the forward injection regime.

From Fig.~3(b), we can also infer that the forward resonant voltage exhibits a lower bound as the silicon concentration approaches the solubility limit [\onlinecite{Faria2012,Lugani2014}]. This trend is consistent with the resonant voltages reported by multiple groups across different GaN/AlN RTD designs. Consequently, achieving forward resonant injection at voltages below 2~volts in polar GaN/AlN double-barrier heterostructures, would require advanced bandgap design techniques, going beyond uniform doping methods.

\section{Space-Charge Effects on the RTD Small-signal Impedance}

\begin{figure}[t]
	\centering
	\hypertarget{Fig-GV}{}
	\includegraphics[width=0.47\textwidth]{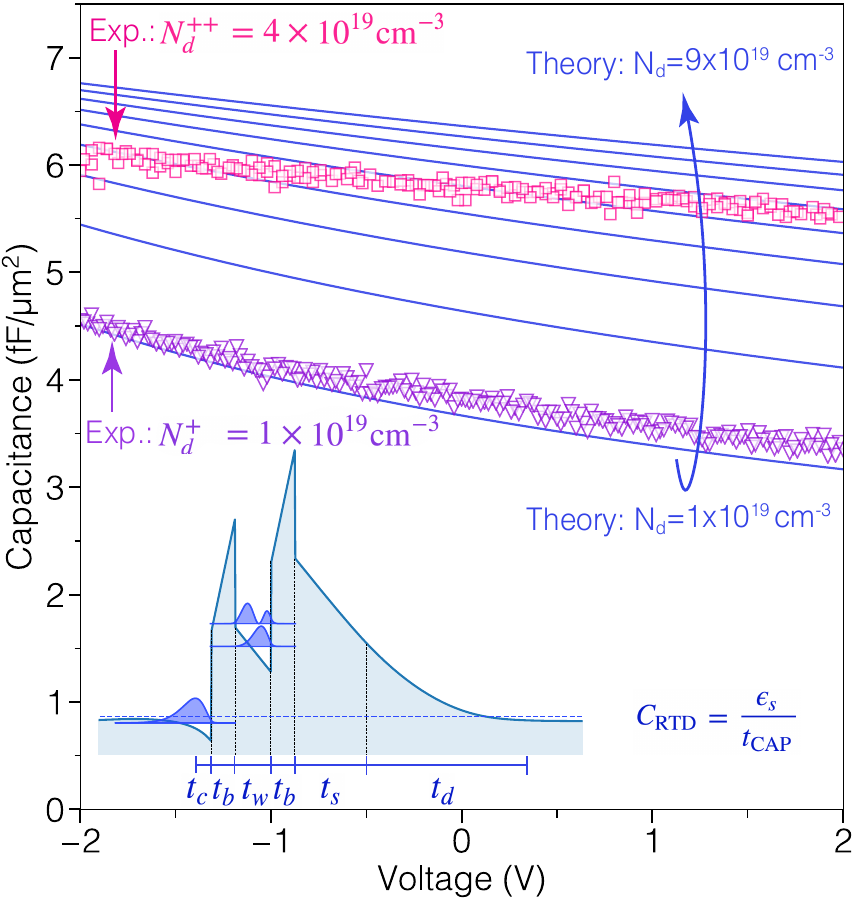}
	\caption{III-nitride RTD capacitance measured under close-to-equilibrium conditions for the two RTD designs with different contact doping concentrations $N_d^+$ and $N_D^{++}$. Employing our analytical model for polar RTDs we also compute the theoretical RTD capacitance, given by Eq. (3), for the structure shown in Fig.~1(a), while varying the doping concentration between $1\times 10^{19}$ and $9\times 10^{19}$~cm$^{-3}$ (blue curves). The agreement between the experimental and theoretical results reveals that polar RTDs behave effectively as parallel-plate capacitors, when the population of the resonant level is negligible. The inset depicts a general polar RTD structure which shows the various layer thicknesses employed to calculate the effective parallel-plate distance $t_{cap}=t_d+t_s+2t_b+t_w+t_c$ and RTD capacitance $C_{RTD}=\epsilon_s/t_{cap}$.}
\end{figure}

As we have seen in the previous section, the interplay between the strong polarization charges and free carriers result in a redistribution of charge that substantially modifies the resonant tunneling transport regime. Effects derived from this phenomenon manifest not only in the dc characteristics of polar RTDs, but also in their frequency response. To determine the influence of the polar active region in the small-signal impedance, we analyze the charging mechanism of the polar double-barrier heterostructure. Under close-to-equilibrium conditions, the quantum well is depopulated and charge is only present within the emitter accumulation well, and distributed across the collector space-charge region [See Fig.~1(b)]. This charge configuration behaves effectively as a parallel-plate capacitor in which the extension of the space-charge region and population of the accumulation subband are modulated when a time-dependent signal is applied. Within this regime, we can use the depletion approximation to write an expression for the charge stored within the depletion layer: $Q_\text{Space-Charge}=qN_dt_d$. Therefore, the small-signal RTD capacitance would be given by:
\begin{equation}
C_\text{RTD}=\frac{dQ_\text{Space-Charge}}{dV_\text{bias}}=qN_d\frac{dt_d}{dV_\text{bias}}.
\end{equation}

The analytical model introduced in Ref. [\onlinecite{Encomendero2017}] can be employed to derive a closed-form formula for the depletion width, $t_d$. By substituting the first equation from the present report into equation (1) of Ref. [\onlinecite{Encomendero2017}], we obtain an expression that relates $t_d$ to the applied bias. After implicit differentiation and using Eq. (2), we can write the RTD capacitance as
\begin{equation}
C_\text{RTD}=\frac{\epsilon_s}{t_d+t_s+2t_b+t_w+t_c}=\frac{\epsilon_s}{t_\text{cap}}.
\end{equation}
Here $t_d$ is the width of the depletion region, $t_s$ is the width of the collector spacer, and $t_c$ is the distance from the centroid of the accumulation subband to the emitter barrier, as depicted in the inset of Fig.~4.

Equation (3) reveals that polar RTDs behave effectively as a parallel-plate capacitor with an effective plate separation given by $t_\text{cap}=t_d+t_s+2t_b+t_w+t_c$. We point out that this simple picture is valid only when the charge inside the quantum well is negligible compared to the population of the accumulation well and collector space-charge. As the bias increases and ground-state population builds up, deriving an analytical expression for the RTD capacitance becomes difficult. In this regime a self-consistent approach would be necessary for computing the charge stored within the active region and the effective RTD capacitance [\onlinecite{Lake2003}]. 

To verify the validity of our model, we measure the small-signal admittance of our devices at room temperature. A 10-mV ac signal is applied to the RTDs while sweeping the dc voltage within the range $-2\,\text{V}\leq V_\text{bias}\leq +2\,\text{V}$. To extract the device capacitance it is required that the susceptance dominates the overall device admittance, thus the phase angle should be approximately 90$^\circ$. After verifying this behavior, the RTD capacitance is extracted employing a small-signal circuit model composed of a capacitor in parallel with a high-impedance resistor.

The measured capacitance-voltage (\textit{C-V}) curves are shown in Fig.~4, which also displays the theoretical capacitance values obtained using equation (3). In contrast to the initial reports of capacitance in III-nitride RTDs [\onlinecite{Kurakin2006}], our devices exhibit repeatable current-voltage characteristics and also stable $\textit{C-V}$ behavior verified by dual-sweep measurements. The capacitance of polar RTDs exhibits a monotonic decrease as the bias is swept from negative to positive values, consistent with the widening of the depletion region. This trend contrasts with the symmetric \textit{C-V} curve measured in nonpolar III-V RTDs [\onlinecite{Wei1993,Tang2001}].

Good quantitative agreement is observed over the whole voltage range for the GaN/AlN RTD with $N_d^+=1\times 10^{19}$~cm$^{-3}$. On the other hand, the slight deviation encountered at high doping densities might be due to a couple of effects related to the RTD growth as well as the approximations considered in our model. During epitaxial growth, it is likely that dopant diffusion results in a doping tail inside the spacer region. Consequently, the smaller effective parallel-plate distance $t_{cap}$, would lead to a slight increase of the experimental RTD capacitance. Since non-unifom doping lies beyond the scope of our analytical model, it would be reasonable to expect a small deviation between the measured and theoretical RTD capacitances. However, the overall good quantitative agreement between theory and experiments validates the proposed parallel-plate model as an appropriate physical picture within the low-bias range.

\section{Concluding Remarks}

To summarize, we have presented a comprehensive study of the dc and ac tunneling transport characteristics of polar GaN/AlN double-barrier heterostructures. The interplay between the fixed polarization charge inside the active structure and mobile charge in the access regions results in a pronounced modulation of the conduction-band profile, which manifests in the resonant tunneling conditions and frequency response. These effects are studied experimentally by systematically varying the doping concentration in the contacts, thereby modulating the width of the collector space-charge layer. As the doping concentration increases, we show an enhanced control over the energies of the quasi-bound states, leading to a lower resonant-tunneling voltage. In addition, due to the narrower thickness of the space-charge region, we observe for the first time repeatable negative differential conductance under both bias polarities of GaN/AlN RTDs, attesting to the high quality of the double-barrier heterostructure. These findings shed light on the important effects of the space-charge regions in the resonant tunneling transport characteristics of polar RTDs.

To explore the consequences of increasingly high doping levels, we also develop a model which predicts a lower bound for the minimum achievable resonant-tunneling voltage as the doping density approaches the solubility limit. Moreover, the effects of the space-charge distribution on the device admittance are also experimentally probed by measuring the small-signal frequency response of GaN/AlN RTDs. These results indicate that under close-to-equilibrium conditions, polar RTDs behave effectively as parallel-plate capacitors. This phenomenon is completely captured by our polar RTD model from which we derive an analytical expression for the bias-dependent RTD capacitance. The good quantitative agreement between the theory and experiments, confirms the validity of our model providing a clear physical picture for the charge storage mechanism exhibited by polar RTDs.

\section{Acknowledgments}

We thankfully acknowledge funding from the Office of Naval Research under the DATE MURI Program (Contract: N00014-11-10721, Program Manager: Dr. Paul Maki) and the National Science Foundation (NSF), under the MRSEC program (DMR-1719875). Partial support from NSF-DMREF (DMR-1534303) and EFRI-NewLAW (EFMA-1741694) programs is also acknowledged. This work was carried out at CNF and CCMR Shared Facilities sponsored by the NSF NNCI program (ECCS-1542081), MRSEC program (DMR-1719875) and MRI DMR-1338010.

\bibliography{z_refs.bib}

\end{document}